# New Failure Rate Model for Iterative Software Development Life Cycle Process

Sangeeta, *Member, IEEE,* Kapil Sharma, *Member, IEEE,* and Manju Bala

*Abstract*—Software reliability models are one of the most generally used mathematical tool for estimation of reliability, failure rate and number of remaining faults in the software. Existing software reliability models are designed to follow waterfall software development life cycle process. These existing models do not take advantage of iterative software development process. In this paper, a new failure rate model centered on iterative software development life cycle process has been developed. It aims to integrate a new modulation factor for incorporating varying needs in each phase of iterative software development process. It comprises imperfect debugging with the possibility of fault introduction and removal of multiple faults in an interval as iterative development of the software proceeds. The proposed model has been validated on twelve iterations of Eclipse software failure dataset and nine iterations of Java Development toolkit (JDT) software failure dataset. Parameter estimation for the proposed model has been done by hybrid Particle Swarm Optimization and Gravitational Search Algorithm. Experimental results in-terms of goodness-of-fit shows that proposed model has outperformed Jelinski Moranda, Shick Wolverton, Goel Okummotto Imperfect debugging, GS Mahapatra, Modified Shick Wolverton in 83.33 % of iterations for eclipse dataset and 77.77% of iterations for JDT dataset.

*Index Terms*—Software development life cycle, Iterative software development life cycle, Optimization, Nature-inspired algorithms, Software reliability models.

## ACRONYMS

| | |
|---|---|
| SDLC | Software Development Life Cycle |
| LLF | Log Likelihood Function |
| MLE | Maximum Likelihood Estimation |
| SSE | Sum of Squared Error |
| MSE | Mean Square Error |
| JM | Jelinski Moranda |
| PSO-GSA | Particle Swarm Optimization and Gravitational Search Algorithm |

Sangeeta is with Delhi Technological University, New Delhi, India (e-mail: sangeeta@dtu.ac.in).
K. Sharma is with Delhi Technological University, New Delhi, India (kapil@ieee.org).
Manju Bala is with Indraprastha College for Women, Delhi University, New Delhi, India (manjugpm@gmail.com )

## NOTATIONS

| | |
|---|---|
| $\lambda(t_i)$ | Failure Intensity |
| $f(t_i)$ | Probability Density Function |
| $F(t_i)$ | Cumulative Distribution Function |
| $R(t_i)$ | Reliability Function |
| $L(N)$ | Likelihood Function |
| $\gamma$ | Modulation factor for representing changing needs in an iteration of software development |
| $\mu$ | Modulation parameter that represents newly added functionality and user acceptance in an iteration |
| $N$ | Number of initial faults in iteration |
| $p$ | Probability of fault removal in iteration |
| $r$ | Probability of fault introduction in iteration |
| $n_{i-1}$ | Cumulative number of failures at $(i-1)^{th}$ failure interval |

## I. INTRODUCTION

WITH the growing advances in the digital world, software development demand from industries is growing at an exponential rate. Due to enormous demand and lack of time and budget, software companies are not able to develop fault-free software. Latest tools and techniques have been applied for the development of defect-free software, but still, it is not possible for software developers to develop defect-free software practically. Software must go through exhaustive testing and debugging, which requires time and money to enhance the reliability [1] [2]. The occurrence of fault is inevitable in the current demand of software. There should have some means to avoid software failures so that devastating losses whether related to life or any other field could be evaded.

According to IEEE standard 729 [3], reliability is the most significant quality aspect of the software. If we could measure the reliability of software under development, better we can predict whether the software would be operational in the future or not. Reliability estimation process must be precise to provide information to the manager like what should be the release time of the software and amount of man-hour consumption etc. while developing any software [4]. Software reliability models are one of the ways to simulate software reliability estimation curve to predict the reliability of the system under study. Numerous reliability estimation models for software have been developed, and all are working on specific applications, specific environments, datasets and assumptions made by them. Then, what is the need for developing new software reliability models?

First need lies in the fact that, among the available research in software reliability model development [5] [6] [7] [8] [9] [10], all developed models are basically made for reliability estimation of software's developed under traditional waterfall SDLC process. Second, research publications are screening no more than 31% experimental researches and among this only 13 % are purely experimental [11]. This low number is due to the reason that public experimental data sets in software reliability estimation are very narrow and producing reliability data through experimentation usually require long time cluster. Third, most generally developed models are NHPP based and very few failure rate based models [12] [13] [14] [15] [16] [17] [18] [19] [20] [21] [22] [*23*].

Failure rate-based models are among the earliest software reliability estimation models used in industrial and academia. These models are grounded on JM software reliability model [9]. They need further enhancement, so that more realistic assumptions like imperfect debugging and factors for exact reliability growth estimation can be incorporated into the model. Further, all software reliability models are developed under waterfall SDLC process. Waterfall SDLC process for the development of software assumes that requirements from the end users are stable and it delivers whole software at one shot in the end [24] [25]. This may generate risks for the users as they do not have information till end what they will get. These limitations construct a need for the use of another methodology while modeling software reliability. Yet now, no research in software reliability assessment has been found which are based on latest SDLC process.

Software reliability is primarily dependent on the process of software development. The software developed using latest SDLC approach can capture and implement all the user requirements within time and budget. Earlier waterfall SDLC process based reliability estimation models are very much stable and found to be unsuitable for the software industries because there are high risk and uncertainty of change in requirements. An iterative SDLC process [26] is the latest methodology of software development and is a practical method of step-wise top-down refinement approach to the software development that replaces the waterfall SDLC process. It gathers user requirements in each iteration of software development and implements software with less uncertainty in risk and user satisfaction. Nowadays the latest software development processes are based on iterative software development process and these are like Rational Unified Process (RUP) [27], Adaptive method [28] [*29*] (Agile methodology), XP (Extreme Programming) [*30*], spiral [31] model and AZ [*32*] development processes. These SDLC processes are considered as an aid of producing reliable software based on iterative model in development. Moreover, these are found to be used practically in industries and academia for software development and promote iterative development process.

Further accurate software reliability estimation is dependent on the selection of optimum parameter values of the models. The techniques for optimizing parameter values of the software reliability models are available, through the various classical methods of parameter estimation [5] [*33*]. These methods are based on number of constraints and may fall in local maxima and do not converge to global maxima in the multimodal cases. An alternative to these classical mathematical optimization methods are nature-inspired optimization algorithms for the solution of non-differential, non-linear and multimodal problems [34] [35] [36].

The proposed failure rate model incorporates iterative SDLC process by replacing earlier waterfall based SDLC process. Further, it assumes imperfect debugging during each of iteration. There is always a possibility of fault introduction with feature addition in each of the iterations of software development. All latest iterative SDLC processes can be used to predict the reliability by applying the proposed failure rate model. Existing failure rate model cannot be applied to the current Software development methodology. The proposed model takes care of complexity and paradigm shift of iterative based software development process by introducing modulation factor.

Rest of the paper is schematized as follows: Section 2 exposes the existing literature of the software reliability models and parameter estimation algorithms used for software reliability modeling. In section 3 a new classification scheme for software reliability models based on iterative SDLC process is discussed. In Section 4 a new software reliability model is proposed. Section 5 discusses the experimental setup and results of the proposed model and the last section concludes the paper.

II. LITERATURE REVIEW

For reliability estimation of the software more than 300 SRGMs have been developed under various categories of reliability estimation models like [6] [*5*] [37] [*38*] [39] [40] [41] [42] [43]. Experimental research work for software reliability estimation is found to be very few, reason being limited failure data and requisite of long time period. From the developed models, no model is well applicable in all type of applications and fits only with a specific application. Some of the model groups are having assumptions that are considered better and have higher predictive quality than other models. Most of the model development comprises NHPP based models and very few are centered on the failure rate behavior. Literature in this paper discusses well-known failure rate behavior-based models.

Jelinski-Moranda model [9] is the first software reliability estimation model. This model is perfect debugging based and assumes that there is fixed, constant and unknown number of initial faults in the software. The time between failures is assumed as independent and exponentially distributed. Shick-Wolverton (SW) model [10] assumed that failure rate function is proportional to the current total number of faults and the time since last failure. This idea has been evolved after modification from the basic JM Model. L.I.A. Turk and E.G. Alsolami developed a model using JM model by applying Weibull distribution function for the amount of debugging time between fault occurrences [44]. Modified Shick-Wolverton model [6] [5] [12] modified the SW model by incorporating a cumulative amount of faults at a time. These models incorporate perfect debugging process while making assumptions for the model development. Goel and Okumotto first time proposed imperfect debugging based model that incorporates imperfect debugging phenomenon by introducing fault removal probability [5]. JM Geometric Model [13] modified JM model by assuming a geometric decrease in program failure rate at failure times. Littlewood and Sofer [14] proposed a model with Bayesian modification in the JM model and provided improvement for estimating the correct reliability of the software. Further modification to the JM model based on the cloud theory has been applied by Luo, Cao, Tang, and Wu [15]. Y.C. Chang and C. Liu proposed a generalized model by extending the JM model using a truncated distribution that matches up a self-exciting point process [45]. Mahapatra and Roy proposed a model and assumed the imperfect debugging behavior of the software with fault removal probability, probability of fault not removed and probability of fault introduction [16]. This model introduced fault introduction probability while removing any fault during debugging.

In failure rate behavior-based models, JM model assumed perfect debugging with single failure between time intervals after that multiple failure were proposed by SW but considers perfect debugging. Imperfect debugging based GOI model consider single failure in a time interval. Littlewood and Luo modified JM model according to the Bayesian and cloud theory respectively. Mahapatra *et al*. introduced probability for fault introduction in the software. In these enhancements to the earliest failure rate model, multiple faults with imperfect debugging have not been covered by any model. These entire model enhancements are made for software's developed under most basic waterfall development life cycle process. There is a need to incorporate latest software development processes for precise estimation of software reliability.

Parameter estimation process plays a crucial role in estimating accurate reliability of software product. There are number of models which are based on the classical methods for parameter estimation [33] [46]. These are numerical technique based methods (like Newtonian method, Gauss-Siedal method, etc.) [47], and these methods are doing well in a number of mathematical optimization problems. Nowadays meta-heuristic based optimization methods are found to be performing well in different domains. The algorithms like Particle Swarm Optimization [48] inspired by the bird flocks and Genetic Algorithm [49] inspired by the natural evolution phenomenon have been well applied in various fields including the field of software reliability estimation. A. Sheta and Jordan Al-Salt [50] applied PSO for SRGM parameter estimation. Malhotra and Negi applied PSO for reliability modeling [51]. Jin and Jin [52] applied improved swarm intelligent approach for parameter optimization of SRGM based on s-shaped testing effort function. Hybrid nature-inspired algorithms are found to be performing well in various domains. Mirjalili and Hashim proposed a hybrid PSO-GSA algorithm by hybridizing PSO and Gravitational Search Algorithm for mathematical function optimization [53]. Abraham *et al*. proposed a hybrid differential and ABC algorithm [54]. Mirjalili and Wang proposed a binary optimization for hybrid PSO-GSA algorithm [55]. As hybrid algorithms are performing well in different fields, these algorithms could be applied in software reliability model parameter estimation so that accurate software reliability could be estimated.

III. CLASSIFICATION OF SOFTWARE RELIABILITY MODELS

This paper classifies software reliability models based on iterative SDLC process. Iterative model is like a cyclic process that mainly focuses on initial very simple implementation that progressively gains complexity and wider set of features till the final system is completed. Most recently built iteration and its feedback from evaluation are used in the next iteration, and accordingly, refinements are made in future iterations. Each of iteration provides enhancements and at-least found to be better than the last. Iterative model adapts rapidly to ever-changing needs of projects and clients within lowest time and budget.

Sharma *et al*. [56] has given classification of software reliability models according to waterfall SDLC process. In this paper, keeping iterative software development practice in mind software reliability models are categorized in Fig. 1. Every phase in iterative software development process is associated with specific requirements and future plans. Fig. 1 shows that each of the iteration is associated with a modulation parameter showing the added functionality and user acceptance level in each iteration. Initially, system is least reliable, and as the number of iterations proceeds, system moves towards refinement and gains reliability. Models are grouped into five categories. Each model category is assigned to a specific phase in iterative software development process for reliability estimation.

IV. PROPOSED MODEL

Keeping new SDLC processes and technologies, a new failure rate model centered on iterative SDLC process has been proposed. The proposed model has derived from the general class of failure rate behavior-based models and exploits iterative behavior of software development process.
In iterative process of software development, entire software is built and delivered to the end user in iterations. The process starts with simple implementation of key samples of a problem and iteratively enhances existing software releases until full system is implemented [57]. At each of iteration release feedback from the iteration is available for next iteration [58] [59] [60]. Feedback is mainly about the functionality and user aspect of the software. At each released stages of software, not only extensions but design modifications may be made. Each of the iteration makes step wise refinements in an effective way to converge to the full implementation of a problem. Main focus of feedback analysis is to find the amount of defects in an iteration that gets injected in the upcoming iteration [61]. Using feedback in each iteration, analysis of changing needs in an up-coming iteration can be made. These changing needs in each iteration are essential to know because these tells information about how much more defects may occur and how much more effort and functionality is needed in making extensions and design modifications to the released iteration.

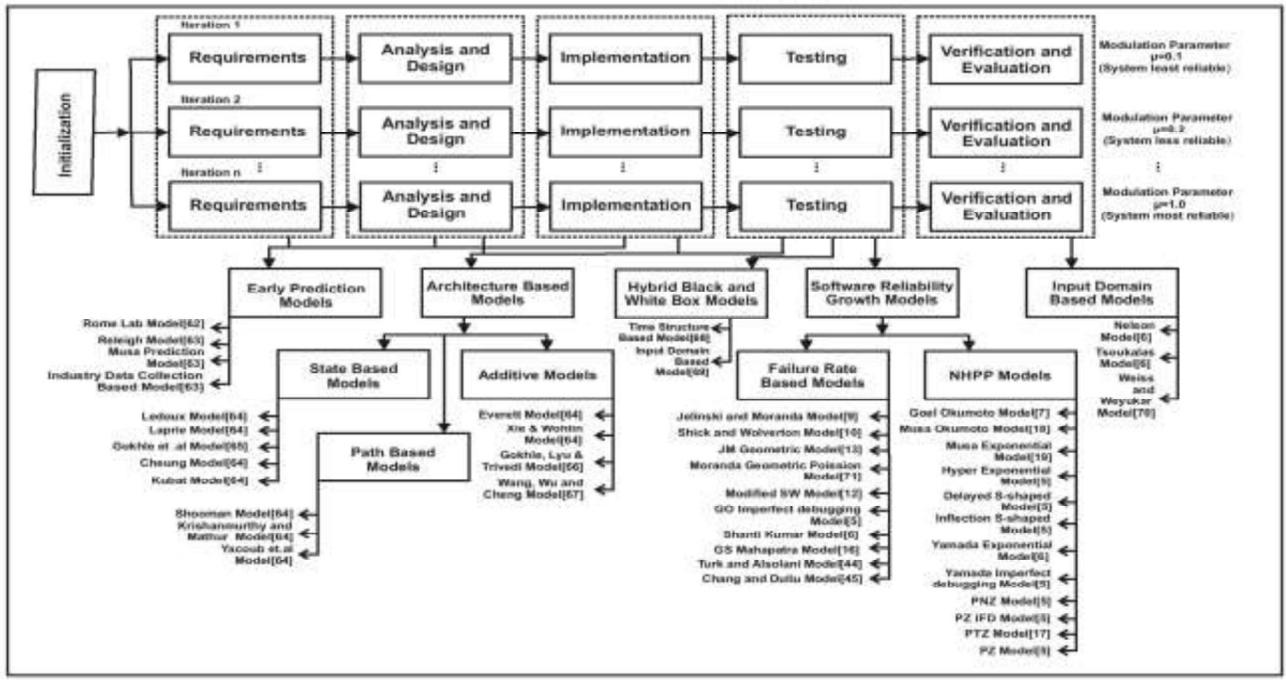

Fig. 1. Classification of Software Reliability Models based on Iterative approach.

In modeling software reliability, there is a need to introduce a factor that will engulf all the changing needs depending on the defect analysis in each of iteration. The varying needs in each of iteration of software development include bug reports, additional functionalities, and testing effort required to find the amount of defects that gets injected / removed in each upcoming iteration. These changing needs are incorporated in proposed failure rate model using modulation factor $\gamma$ given in (1), so that precise reliability growth of the system could be estimated. The proposed model assumes that fault removal process is imperfect. Due to imperfect debugging regenerated faults are induced in successive iterations.

A. *Proposed Model Assumptions*

i. Initial software fault is unknown and constant in an iteration.
ii. Each fault in an iteration is independent and it may be equally likely to cause a failure while testing.
iii. The interval of time between fault occurrences in each iteration is independent and follows an exponential distribution.
iv. Software failure rate remains constant over the intervals between fault occurrences.
v. The software failure rate is proportional to the number of faults that remain in the software and modulation factor $\gamma$.
vi. In each iteration the detected fault is removed with a probability $p$, not removed perfectly with a probability $q$ and new fault may be introduced with a probability $r$. Here $p+q+r=1$ and the probability $p>r$.

Faults are injected from previous iterations along with newly introduced faults in the current iteration. Newly induced faults are caused by added and modified functionalities in a respective iteration of software development. Depending on the number of initial iterative faults, there is a need to modify the amount of resources allocated for debugging in each iteration. The modulation factor $\gamma$ as defined in (1) reflects the modified needs that integrate iterative development processes in software failure rate models. Modulation factor changes its value according to the modulation parameter $\mu$ as shown in (1). Moreover, changing needs in each iteration is different and vary according to (1).

$$\gamma = \mu + (1-\mu)/\mu, \quad (0 < \mu \le 1.0) \qquad (1)$$

Here, $\mu$ is the modulation parameter that represents newly added functionality and user acceptance in the current iteration. Its value is almost 0 at the beginning and becomes 1.0 in the final iteration of iterative software development process. Modulation parameter $\mu$ takes its value from 0 to 1.0 by assuming that with growing number of iterations, level of user acceptance increases from lower to higher. It shows abrupt changes in its value in initial iterations due to preliminary design changes, testing effort, and user acceptance. When it's values reaches near to one then software under development is assumed to be reliable enough and has achieved all the required functionalities to fulfill the end user needs.

## B. Model Formulation

Failure rate function ($\lambda(t_i)$) with imperfect debugging is modeled in (2).

$$\lambda(t_i) = \phi[N - \frac{(n_{i-1})(\gamma)}{(i-1)}(p-r)], \quad i-1,2....N \quad (2)$$

Where
- $\gamma$  Modulation factor for representing changing needs in each of the iteration of software development
- $n_{i-1}$  Cumulative number of failures at $(i-1)^{th}$ failure interval
- $N$  Number of initial faults in software
- $\phi$  Proportionality constant

Cumulative Density Function $F(t_i)$ and Reliability Function $R(t_i)$ is calculated in (3) and (4)

$$F(t_i) = 1 - e^{[-\phi[N - \frac{(n_{i-1})\gamma}{i-1}(p-r)]t_i]} \quad (3)$$

$$R(t_i) = e^{[-\phi[N - \frac{(n_{i-1})\gamma}{i-1}(p-r)]t_i]} \quad (4)$$

When $p = 1, r = 0$ and $\gamma$ varies as in (5), proposed model behaves as JM model.

$$1, \frac{2i}{i+1}, \frac{3i}{i+4}, \frac{4i}{i+8}, \frac{5i}{i+13} .... \quad (5)$$

Following a variation of $\gamma$ in (5) and considering $p$ being the probability of fault removal and $r$ as the fault introduction probability then model behaves as the GS Mahapatra *et al.* model [16].

## C. Parameter Estimation

In the proposed model there are three unknown parameters $N, n$, and $\phi$ these parameters are estimated at different values of the $\gamma$. MLE has been used to estimate the values of the parameters. Parameter estimation by MLE method requires solutions of complex equations by maximizing the likelihood of model parameters. Probability density function $f(t_i)$ for the proposed model is given in (6).

$$f(t_i) = \phi[N - \frac{(n_{i-1})\gamma}{i-1}(p-r)] \cdot e^{[-\phi[N - \frac{(n_{i-1})\gamma}{i-1}(p-r)]t_i]} \quad (6)$$

The likelihood function $L(N)$ is calculated in (7) using (6).
$L(N) = \prod f(t_i)$

$$= \prod_{i=1}^{n} [\phi[N - \frac{(n_{i-1})\gamma}{i-1}(p-r)]e^{[-\phi \sum_{i=1}^{n}[N - \frac{(n_{i-1})\gamma}{i-1}(p-r)]t_i]} \quad (7)$$

Taking the log of $L(N)$, LLF is calculated in (8).

$$LLF = \ln L(N) = n \ln \phi + \sum_{i=1}^{n} \ln[N - \frac{(n_{i-1})\gamma}{i-1}(p-r)]$$

$$- \sum_{i=1}^{n} \phi[N - \frac{(n_{i-1})\gamma}{i-1}(p-r)] t_i \quad (8)$$

Solution using log likelihood function for parameter estimation involves calculation of its partial derivatives with respect to $N, n$ and $\phi$ respectively and then equating them to value zero. MLE of the parameters are calculated from (9), (10) and (11).

$$\frac{n}{\phi} = \sum_{i=1}^{n} [N - \frac{(n_{i-1})\gamma}{i-1}(p-r)] t_i \quad (9)$$

$$n = \phi \sum_{i=1}^{n} [N - \frac{(n_{i-1})\gamma}{i-1}(p-r)] t_i \quad (10)$$

$$\frac{1}{\sum_{i=1}^{n}[N - \frac{(n_{i-1})\gamma}{i-1}(p-r)]} = \phi \sum_{i=1}^{n} t_i \quad (11)$$

## V. EXPERIMENTAL SETUP AND RESULTS

### A. Application Datasets used for Experimentation

Suitability of the proposed model has been tested using Tera Promise repository bug report files of Eclipse (DS1) and JDT (DS2) open source software at https://zenodo.org/record/268486#.W-QsPpMzY2w. These data sets have been given by An Ngoc Lam. The bug reports contains table having contents as bug_id, summary, description, report time, report time status, commit, and commit time files. Datasets have been extracted from these bug reports and reformatted in time domain format. DS1 includes eight minor releases and four major releases starting from the year 2001 to the year 2013. DS2 includes 3 major releases and 6 minor releases starting from the year 2002 to the year 2014.

Model parameters include $N, n, \phi, \gamma, p$ and $r$. Fault removal probability $p$ and fault introduction probability $r$ cannot be estimated from DS1 and DS2 and depends on the project type and skill set of persons involved in development and testing. Based on these two factors fault removed during testing is assumed 95 % and fault introduced is assumed 3%. Parameters $N, n, \phi, \gamma$ are estimated using hybrid PSO-GSA algorithm and MLE technique in (6), (7) and (8) to maximize the log-likelihood function value using (9), (10) and (11). The goodness of fit for the proposed model is measured using SSE and MSE for each application datasets [56]. These statistics are used for comparison of the proposed model with existing failure rate models.

The values of $\mu$ with respect to iteration are different for different software projects and depend on the type of project and user acceptance levels. Fig. 2 depicts variation in μ values with respect to number of iterations for DS1. The value of $\mu$ decreases in successive iterations 1.0 to 2.0, 2.1 to 3.0 and 3.3 to 3.4 by difference of 0.064, 0.081 and 0.045 respectively. These little deeps in $\mu$ values represent variation in compliance of functionality requirements and lower values of user acceptance levels for eclipse software dataset (DS1). Overall values of $\mu$ shows increasing trends with successive iteration as shown in Fig. 2 and finally it reaches near to 1.0 in final iteration release.

Table 1 SUMMARY OF FAILURE RATE BASED SOFTWARE RELIABILITY MODELS

| Sr. No. | Model Name | Failure Intensity |
|---|---|---|
| 1 | Jelinski-Moranda Model (JM) [9] | $\lambda(t_i) = \phi[N-(i-1)]$ |
| 2 | Schick &Wolverton Model (SW) [10] | $\lambda(t_i) = \phi[N-(i-1)]t_i$ |
| 3 | Goel Okumotto imperfect debugging Model(GOI) [5] | $\lambda(t_i) = \phi[N - p(i-1)]$ |
| 4 | G.S.Mahapatra et al. Model [16] | $\lambda(t_i) = \phi[N - p(i-1) + r(i-1)]$ |
| 5 | Modified S-W Model(MSW) [5] | $\lambda(t_i) = \phi[N - (n_{i-1})]t_i$ |
| 6 | Proposed Model | $\lambda(t_i) = \phi[N - \frac{(n_{i-1})(\gamma)}{(i-1)}(p-r)]$ |

Fig. 3 is revealing the change in functionalities and its corresponding increase in user acceptance for DS2 of JDT software product. At each of the iteration release there is increase in user acceptance level and it is illustrated with the values of $\mu$ in all successive iterations. For major iteration releases the value of $\mu$ are 0.0869, 0.1108 and 0.2144, for minor iteration these values are 0.1381, 0.2684, 0.3677, 0.6787, 0.7241 and 0.9539. The user acceptance increases in all iterations but a large variation of 0.31102 in user acceptance level is found in 3.4 to 3.5 iteration. It represents enhanced functionality and accomplishment of all requirements at end-users. Overall values of $\mu$ in successive iterations are reflecting variation in functionalities and user acceptance level in Fig. 3 that increases by small amount in each iteration and finally reaches near to 1.0 in last iteration release.

### B. Result Analysis

#### 1) CASE 1: Eclipse Software Dataset (DS1)

In this section, the goodness-of-fit of the proposed model is calculated and compared it with existing models given in Table 1 for DS1. The value of $\gamma$ for major iterative release 1.0, 2.0, 3.0 and 4.1 are 2.2666, 2.8763, 3.7891 and 1.1258 respectively. In minor iteration release, the value of $\gamma$ for iterative release 2.1, 3.1, 3.2, 3.3, 3.4, 3.5, 3.6 and 4.2 are 2.6343, 3.2217, 2.5788, 2.3121, 2.7295, 2.6434, 2.0435 and 1.00009 respectively.

The value of $\gamma$ shows large changes in successive major iterations as compared to successive minor iterations due to varying needs in major and minor iterations. The goodness of fit of the models is shown in Table 2. From the analysis of results, it is found that proposed model fits well in term of SSE in all iterations of DS1 except at iteration number 3.1 where GOI model, GS Mahapatra model, and SW model outperforms proposed model.

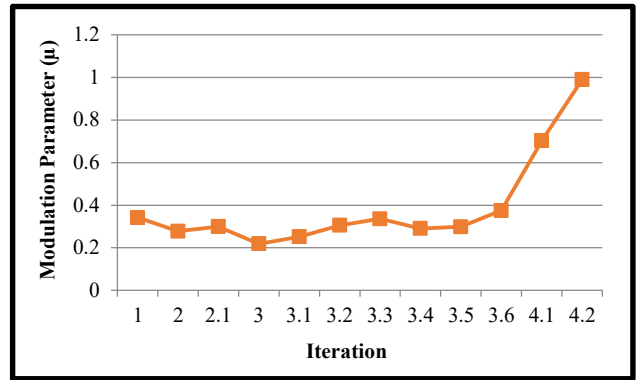

Fig. 2 Plot of μ versus Iteration for DS1

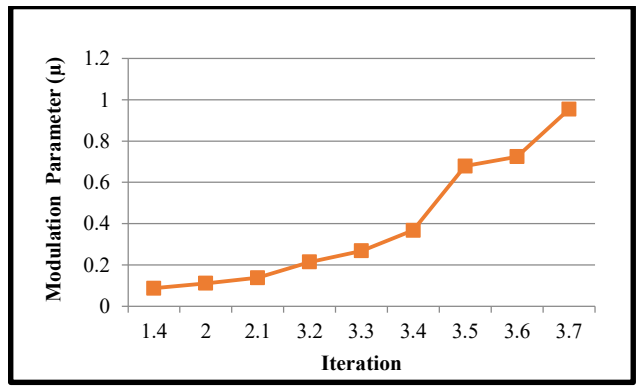

Fig. 3 Plot of μ versus Iteration for DS2

The proposed model has clear-cut outperform all models under comparison in 11 iterations for DS1. It shows the lowest values of SSE in 91.6 % iterations. In term of MSE proposed model is winner in nine iterations. In iteration 2.1 MSW model is performing well than the proposed model. In iteration 3.3, GOI model, SW model, and GS Mahapatra model are performing well than the proposed model. There is a tie among proposed model and GS Mahapatra model in iteration 2.0, where both of them outperform all other models. The proposed model has clear-cut outperform other models in 75 % iterations by achieving lowest value of MSE. Result shows that proposed model has given a significantly better fit to iterative data by adapting according to varying needs of different iterations.

The evaluation and comparison of goodness-of-fit of all models in Table 1 in terms of SSE and MSE for all iterations shows that the proposed model has promising technical merit in the sense that it provides development terms with both iterative SDLC requirements and traditional reliability measures.

#### 2) CASE 2: JDT Dataset

To test the applicability of the proposed model in this case, dataset DS2 is used. Table 3 is showing the estimated values of model parameters and goodness of fit criteria. The value of $\gamma$ for major iterative release 1.4, 2.0 and 3.2 are 10.5954, 8.1358 and 3.8778 respectively. In minor iteration releases the value of $\gamma$ for iterative release 2.1, 3.3, 3.4, 3.5, 3.6 and 3.7 are 6.3815, 2.9946, 2.0872, 1.1521, 1.1052 and 1.0022, respectively. These values of $\gamma$ are showing large deviations in successive major releases as compared to successive minor releases due to the varying needs in each of the major and minor iterations. Table 3 is showing the goodness-of-fit measures of all models depicted in Table 1 for DS2. Out of total nine iterations in DS2, proposed model outperform other models under comparison in eight iterations, in terms of SSE values. JM model performs well than proposed model in iteration 3.4. Proposed model attains lower-most value of SSE in 88.8 % of iterations. In terms of MSE values for nine iterations proposed model out-performs other models in six iterations. In iteration 2.0, GS Mahapatra model and GOI model are performing better than the proposed model. In iteration 3.7, proposed model outperforms JM, GOI, SW and MSW models except GS Mahapatra model. The proposed model outperforms GOI, SW, GS Mahapatra and MSW model except JM model in iteration number 3.6. The proposed model has performed better than all other models under comparison in 66.6% iterations by attaining the minimum value of MSE. Result analysis in Table 3 shows that the proposed model has significantly better fit to iterative data by fine-tuning to varying needs of different iterations.

Table 2 GOODNESS-OF-FIT ES/TIMATED USING DS1 (ECLIPSE SOFTWARE FAILURE DATASET)

| Sr. No. | Model | Iteration1.0 | | | Iteration2.0 | | | Iteration2.1 | | | Iteration 3.0 | | |
|---|---|---|---|---|---|---|---|---|---|---|---|---|---|
| | | Estimated Parameter values | SSE | MSE | Estimated Parameter values | SSE | MSE | Estimated Parameter values | SSE | MSE | Estimated Parameter values | SSE | MSE |
| 1 | JM Model | $\Phi$=5.74E-06, N=4 | 5.23 | 5 | $\Phi$=2.68E-05, N=8 | 375.51 | 17.05 | $\Phi$=5.039E-06, N=27 | 452.2 | 16.74 | $\Phi$=2.40E-06, N=124 | 56948 | 605.82 |
| 2 | GOI Model | $\Phi$=1.64E-05, N=3 | 1.51 | 1 | $\Phi$=2.89E-06, N=28 | 92.01 | 4.18 | $\Phi$=1.95E-06, N=26 | 133.34 | 4.93 | $\Phi$=1.37 E-06, N=100 | 45510.1 | 559.22 |
| 3 | SW Model | $\Phi$=2.124E-06, N=3 | 17.31 | 17.00 | $\Phi$=4.70E-06, N=28 | 218.81 | 9.91 | $\Phi$=3.83E-06, N=29 | 126.6 | 4.66 | $\Phi$=2.92 E-05, N=121 | 39981 | 494.69 |
| 4 | GS Mahapatra Model | $\Phi$=7.77E-07, N=3 | 1.26 | 1 | $\Phi$=2.53 E-05, N=23 | 55.46 | 2.5 | $\Phi$=3.20E-06, N=28 | 131.31 | 4.85 | $\Phi$=1.25 E-06, N=125 | 55122.2 | 586.40 |
| 5 | MSW Model | $\Phi$= 2.23E-05, N=2, n=58 | 5.0 | 4.9 | $\Phi$=1.52E-06, N=9, n=209 | 528.7 | 15.89 | $\Phi$=1.36E-05, N=18, n=316 | 118.67 | 3.89 | $\Phi$=1.29E-06, N=123, n=298 | 66890.4 | 679.32 |
| 6 | Proposed Model | $\Phi$= 2.86E-05, N=5, n=11, $\Upsilon$=2.2666, | 1.24 | 0.66 | $\Phi$=4.70E-06, N=27, n=362, $\Upsilon$=2.8763, | 45.12 | 2.5 | $\Phi$=1.67E-06, N=30,n=415, $\Upsilon$=2.6343, | 101.02 | 4.39 | $\Phi$=2.98E-05, N=102,n=5690, $\Upsilon$=3.7891 | 39705.2 | 484.14 |

Table 2 continued…

| Sr. No. | Model | Iteration3.1 | | | Iteration3.2 | | | Iteration3.3 | | | Iteration 3.4 | | |
|---|---|---|---|---|---|---|---|---|---|---|---|---|---|
| | | Estimated Parameter values | SSE | MSE | Estimated Parameter values | SSE | MSE | Estimated Parameter values | SSE | MSE | Estimated Parameter values | SSE | MSE |
| 1 | JM Model | $\Phi$=2.99 E-05, N=100 | 6091.1 | 45.45 | $\Phi$=2.21E-06, N=122 | 13940.73 | 119.14 | $\Phi$=2.95E-05, N=117 | 6810.14 | 58.24 | $\Phi$=8.28E-06, N=53 | 1153.32 | 23.53 |
| 2 | GOI Model | $\Phi$=3.07E-07, N=126 | 1018.8 | 7.59 | $\Phi$=8.61E-06, N=106 | 7246.05 | 61.93 | $\Phi$=1.25E-06, N=115 | 9371.1 | 80.09 | $\Phi$=3.06E-06, N=50 | 1545.41 | 1320.9 |
| 3 | SW Model | $\Phi$=2.99E-05, N=125 | 1123 | 8.38 | $\Phi$=5.40E-06, N=90 | 12585 | 107.56 | $\Phi$=2.95 E-05, N=123 | 7318.81 | 62.55 | $\Phi$=2.93 E-05, N=54 | 1153.73 | 23.53 |
| 4 | GS Mahapatra Model | $\Phi$=6.62E-06, N=130 | 1015 | 7.57 | $\Phi$=1.25E-06, N=115 | 7794.28 | 68.97 | $\Phi$=3.06E-06, N=114 | 2217.4 | 18.99 | $\Phi$=1.63E-05, N=55 | 905.4 | 18.46 |
| 5 | MSW Model | $\Phi$=1.36E-05, N=135, n=4979 | 2000.7 | 15.9 | $\Phi$=2.08E-06, N=132, n=4108 | 9685.12 | 89.67 | $\Phi$=5.31E-06, N=121,n=3129 | 2903.9 | 27.87 | $\Phi$=1.72E-05, N=52, n=2094 | 3589.96 | 29.79 |
| 6 | Proposed Model | $\Phi$=2.96E-05, N=136,n=6132, $\Upsilon$=3.2217 | 1589.82 | 12.22 | $\Phi$=2.98 E-05 , N=122,n=6910, $\Upsilon$=2.5788 | 952.2 | 8.14 | $\Phi$=2.96E-05, N=123,n=7879, $\Upsilon$=2.3121 | 2203.05 | 19.49 | $\Phi$=2.92 E-05, N=53, n=1411, $\Upsilon$=2.7295 | 520.23 | 11.55 |

Table 2 continued…

| Sr. No. | Model | Iteration 3.5 Parameter estimated values | SSE | MSE | Iteration 3.6 Parameter estimated values | SSE | MSE | Iteration 4.1 Parameter estimated values | SSE | MSE | Iteration 4.2 Parameter estimated values | SSE | MSE |
|---|---|---|---|---|---|---|---|---|---|---|---|---|---|
| 1 | JM Model | $\Phi=1.63E-06$, N=30 | 373.3 | 15.54 | $\Phi=6.24E-07$, N=28 | 199.91 | 7.65 | $\Phi=2.20E-05$, N=19 | 286.65 | 22 | $\Phi=1.92E-05$, N=31 | 441.12 | 15.21 |
| 2 | GOI Model | $\Phi=3.81E-06$, N=29 | 137.74 | 5.71 | $\Phi=3.208E-06$, N=28 | 160.01 | 6.15 | $\Phi=1.58E-05$, N=17 | 111.21 | 8.54 | $\Phi=3.20E-06$, N=30 | 121.21 | 4.17 |
| 3 | SW Model | $\Phi=1.70E-07$, N=29 | 551.1 | 22.96 | $\Phi=3.74E-06$, N=27 | 154.54 | 5.92 | $\Phi=1.93E-05$, N=19 | 284.86 | 22.1 | $\Phi=6.50E-06$, N=31 | 193.34 | 6.66 |
| 4 | GS Mahapatra Model | $\Phi=3.81E-06$, N=29 | 137.8 | 5.71 | $\Phi=6.65E-07$, N=28 | 152.23 | 5.86 | $\Phi=5.53E-06$, N=17 | 129 | 9.92 | $\Phi=7.29E-07$, N=31 | 166.65 | 5.72 |
| 5 | MSW Model | $\Phi=1.26E-05$, N=17, n=2663 | 489.74 | 18.24 | $\Phi=2.97E-05$, N=17, n=2624 | 315.00 | 14.92 | $\Phi=2.68E-05$, N=8, 1529 | 178.22 | 7.34 | $\Phi=2.97E-05$, N=30, n=3729 | 704.86 | 26.87 |
| 6 | Proposed Model | $\Phi=2.82E-05$, N=25, n=416, $\Upsilon=2.6434$ | 86.62 | 4.3 | $\Phi=2.94E-05$, N=30, n=401, $\Upsilon=2.0435$ | 91.15 | 4.14 | $\Phi=2.87E-05$, N=16, n=136, $\Upsilon=1.1258$ | 49.95 | 5.44 | $\Phi=2.82E-05$, N=32, n=513, $\Upsilon=1.00009$ | 72.24 | 2.88 |

Table 3 GOODNESS-OF-FIT ESTIMATED USING DS2 (JDT SOFTWARE FAILURE DATASET)

| Sr. No. | Model | Iteration 1.4 Estimated Parameter values | SSE | MSE | Iteration 2.0 Estimated Parameter values | SSE | MSE | Iteration 2.1 Estimated Parameter values | SSE | MSE | Iteration 3.2 Estimated Parameter values | SSE | MSE |
|---|---|---|---|---|---|---|---|---|---|---|---|---|---|
| 1 | JM Model | $\Phi=1.46E-05$, N=22 | 1510 | 62.91 | $\Phi=5.63E-08$, N=13 | 1792.56 | 66.37 | $\Phi=2.81E-05$, N=37 | 2638.12 | 203.02 | $\Phi=2.35E-06$, N=50 | 179.79 | 11.19 |
| 2 | GOI Model | $\Phi=4.11E-06$, N=15 | 1517 | 75.85 | $\Phi=4.80E-08$, N=18 | 1486.81 | 55.04 | $\Phi=5.53E-06$, N=35 | 1650.04 | 127.90 | $\Phi=8.16E-07$, N=56 | 317.67 | 5.56 |
| 3 | SW Model | $\Phi=8.18E-07$, N=9 | 2165 | 90.21 | $\Phi=6.99E-06$, N=17 | 2427.45 | 187.9 | $\Phi=5.36E-06$, N=38 | 3278.45 | 252.87 | $\Phi=2.17E-06$, N=58 | 1056 | 15.68 |
| 4 | GS Mahapatra Model | $\Phi=9.00E-06$, N=12 | 1871 | 77.96 | $\Phi=1.08E-06$, N=16 | 1434.79 | 53.11 | $\Phi=1.21E-06$, N=57 | 3808.85 | 293.48 | $\Phi=1.21E-06$, N=57 | 56500.09 | 991.23 |
| 5 | MSW Model | $\Phi=7.43E-06$, N=19, n=242.98 | 2909.88 | 121.09 | $\Phi=1.62E-05$, N=15, n=282.93 | 1870.67 | 69.26 | $\Phi=8.73E-06$, N=29, n=2569.9 | 1741.90 | 164.48 | $\Phi=2.91E-05$, N=52, n=1723 | 2989.06 | 29.69 |
| 6 | Proposed Model | $\Phi=2.99E-05$, N=13, n=160, $\Upsilon=10.5954$ | 1379 | 57.54 | $\Phi=2.97E-05$, N=17, n=199, $\Upsilon=8.1358$ | 1292 | 56.34 | $\Phi=2.82E-05$, N=37, n=234, $\Upsilon=6.3815$ | 683.70 | 75.88 | $\Phi=2.91E-05$, N=60, n=1788, $\Upsilon=3.8778$ | 171.89 | 3.226 |

Table 3 continued…

| Sr. No. | Model | Iteration 3.3 Estimated Parameter values | SSE | MSE | Iteration 3.4 Estimated Parameter values | SSE | MSE | Iteration 3.5 Estimated Parameter values | SSE | MSE | Iteration 3.6 Estimated Parameter values | SSE | MSE |
|---|---|---|---|---|---|---|---|---|---|---|---|---|---|
| 1 | JM Model | $\Phi=2.78E-05$, N=9 | 404 | 33.666 | $\Phi=8.86E-06$, N=18 | 37 | 12.33 | $\Phi=2.49E-05$, N=8 | 57 | 4.384 | $\Phi=2.93E-05$, N=4 | 76 | 7.6 |
| 2 | GOI Model | $\Phi=1.48E-05$, N=16 | 989.9 | 61.825 | $\Phi=2.35E-06$, N=18 | 206.56 | 17.16 | $\Phi=1.06E-06$, N=13 | 76.53 | 25.33 | $\Phi=2.01E-05$, N=11 | 46721 | 359.39 |
| 3 | SW Model | $\Phi=4.70E-06$, N=8 | 589.1 | 40.78 | $\Phi=3.83E-06$, N=16 | 750.9 | 68.00 | $\Phi=2.92E-05$, N=10 | 3019.9 | 119.05 | $\Phi=2.99E-05$, N=29 | 12089 | 367.83 |
| 4 | GS Mahapatra Model | $\Phi=8.92E-06$, N=26 | 604.01 | 37.75 | $\Phi=1.03E-05$, N=28 | 1241.01 | 103.41 | $\Phi=4.78E-07$, N=5 | 38225 | 318.54 | $\Phi=1.26E-05$, N=85 | 11534 | 384.48 |
| 5 | MSW Model | $\Phi=2.99E-05$, N=9, n=1028 | 486.32 | 33.960 | $\Phi=2.91E-05$, N=13, n=1124 | 1839.56 | 146.92 | $\Phi=2.31E-05$, N=3, n=137 | 20687 | 156.94 | $\Phi=1.34E-05$, N=15, n=134 | 34834.09 | 329.21 |
| 6 | Proposed Model | $\Phi=2.97E-05$, N=19, n=179, $\Upsilon=2.9946$ | 31 | 2.583 | $\Phi=2.92E-05$, N=18, n=119, $\Upsilon=2.0871$ | 91 | 11.37 | $\Phi=2.96E-05$, N=7, n=25, $\Upsilon=1.1521$ | 18 | -24.002 | $\Phi=2.67E-05$, N=18, n=171, $\Upsilon=1.1051$ | 192 | 21.33 |

Table 3 continued…

| Iteration 3.7 | | | | |
|---|---|---|---|---|
| Sr. No. | Model | Estimated Parameter values | SSE | MSE |
| 1 | JM Model | Φ=1.50E-06, N=16 | 280 | 45.02 |
| 2 | GOI Model | Φ=1.58E-06, N=13 | 90.011 | 9.01 |
| 3 | SW Model | Φ=5.40E-06, N=26 | 200.9 | 11.08 |
| 4 | GS Mahapatra Model | Φ=5.30E-06, N=12 | 952 | 8.14 |
| 5 | MSW Model | Φ=2.69E-05, N=14, n=1599 | 1109.23 | 57.34 |
| 6 | Proposed Model | Φ=2.74E-05, N=15, n=78, ϒ=1.0022 | 54 | 9.00 |

## VI. CONCLUSION

Failure rate models available in literature are centered on most traditional waterfall SDLC process. However, new software development processes have been developed and found to be more beneficial than waterfall SDLC process, like iterative life cycle processes. Keeping in view, new software development environments and technologies, a new failure rate model by exploiting iterative behavior of software development process is proposed. The changing needs in each of the iteration are reflected in the proposed model using a modulation factor. Calculated values of $\gamma$ parameter are significantly reflecting all changing requirements for each upcoming iterations numerically. These values are meaningfully representing how much impact is of adding and removing new features with the level of user acceptance in each of upcoming iteration. In order to compare the performance of the proposed model, five well-known software failure rate models JM, GOI, SW, GS Mahapatra and MSW have been applied for dataset DS1 and DS2. The proposed model is a clear-cut winner in 11 iterations in SSE and 9 iterations in MSE out of 12 iterations and 8 iteration in SSE and 6 iterations in MSE out of 9 iterations for DS1 and DS2 respectively. Overall, in 83.33 % of iterations for DS1 and 77.77 % of iterations for DS2, the proposed model has shown better results in terms of goodness of fit by successfully incorporating varying needs in each of iteration.

The data collected from real applications and comparison of goodness of fit shows that the proposed model successfully incorporated varying needs in each of iteration and performed better than JM, GOI, SW, GS Mahapatra and MSW models.